**Charge accumulation by Direct Magnetoelectric Effect in ScAlN/Ni Nanoscale Devices**

*Federica Luciano, Emma Van Meirvenne, Ephraim Spindler, Philipp Pirro, Bart Sorée, Mathias Weiler, Stefan De Gendt, Florin Ciubotaru, Christoph Adelmann\**

F. Luciano, E. Van Meirvenne, B. Sorée, S. De Gendt, F. Ciubotaru, C. Adelmann
Imec, 3001 Leuven, Belgium
E-mail: Christoph.Adelmann@imec.be

F. Luciano, S. De Gendt
KU Leuven, Department of Chemistry, 3001 Leuven, Belgium

E. Van Meirvenne, B. Sorée
KU Leuven, Department of Electrical Engineering (ESAT), 3001 Leuven, Belgium

E. Spindler, P. Pirro, M. Weiler
Landesforschungszentrum OPTIMAS & Rheinland-Pfälzische Technische Universität Kaiserslautern-Landau, 67663 Kaiserslautern, Germany

Funding: Research Foundation Flanders (Fonds Wetenschappelijk Onderzoek, FWO) – grant numbers: 1183722N (FL) and 1SH4Q24N (EVM)

This work investigates the direct magnetoelectric effect in thin-film lab scale composite heterostructures comprising a 100 nm thick piezoelectric $Sc_{0.4}Al_{0.6}N$ (ScAlN) and a magnetostrictive Ni with 100 - 200 nm thickness, fabricated on $Si/SiO_2$ substrates. The films are patterned into square pillar arrays with lateral dimensions down to 500 nm x 500 nm. Vibrating sample magnetometry (VSM) measurements reveal in-plane magnetic anisotropy in the Ni films, attributed to strain induced by the underlying ScAlN layer. Nitrogen-vacancy (NV) magnetometry imaging confirms the formation of magnetic domains at remanence in polycrystalline Ni when patterned in sub-microscale structures. Capacitance measurements reveal a ScAlN dielectric constant at the device level consistent with unpatterned thin films, confirming the preservation of electrical integrity at the sub-microscale. The direct magnetoelectric effect is demonstrated through quasi-static charge measurements under applied out-of-plane DC magnetic fields, yielding equivalent open-circuit voltages up to $1.17 \pm 0.18$ mV.

# 1. Introduction

Composite multiferroic heterostructures, comprising piezoelectric and magnetostrictive layers, have attracted considerable attention due to their potential to enable a strong magnetoelectric (ME) effect at room temperature.[1–4] In the so formed heterostructure, the ME coupling is given by the strain transferred from the piezoelectric to the magnetostrictive layer (converse magnetoelectric effect), or vice-versa from the magnetostrictive to the piezoelectric layer (direct magnetoelectric effect). In these strain-mediated systems, an electric field applied to the piezoelectric component can reorient the magnetization of the magnetostrictive layer, while a magnetic field can induce a polarization, or measurable charge, in the piezoelectric layer, respectively. These functionalities hold promise for the development of ultra-low-power spintronic and non-volatile memory devices.[5,6]

Despite the promises of the magnetoelectric effect, realizing sufficiently strong coupling in thin film heterostructures on a rigid substrate, like silicon (Si) has proven challenging due to the clamping effect.[7,8] This effect is a mechanical constraint due to the rigid substrate, which severely reduces the in-plane strain transferred between the magnetostrictive and piezoelectric layer.[9–11] In this configuration the material deposited on the rigid substrate cannot freely expand or contract, this leads to an incomplete strain transfer and the reduction of the magnetoelectric coupling. Therefore, in order to achieve a significant coupling, previous research has focused on different approaches, which includes using one of the constituent phases (either ferroelectric or magnetic) as the substrate in heterostructure fabrication,[12–17] integrating the thin-film magnetoelectric composite onto flexible substrates or membranes,[18–20] or fabricating magnetostrictive nanopillars in a piezoelectric matrix to maximize the interface area.[9,21–23]

While these approaches have shown potential in enhancing the magnetoelectric coupling by mitigating substrate clamping, they often come at the expense of films and devices scalability, and compatibility with standard CMOS processes. A less explored solution to overcome this clamping without forsaking planar integration is geometric scaling. Finite-element simulations predicted that, when the magnetoelectric pillars are scaled down to the submicron range, edge relaxation can locally relieve the clamping effect, enabling significant strain-mediated coupling even on rigid substrates.[24]

In this work, we present the first experimental demonstration of the quasi-static direct magnetoelectric effect in nanoscale devices fabricated on a silicon substrate. Our thin film heterostructure consist of ScAlN as piezoelectric layer and Ni as magnetostrictive layer. By patterning the ScAlN/Ni in pillars with lateral dimension down to 500 nm x 500 nm, we partially overcome the clamping limitations of the rigid substrate and obtain a measurable strain-mediated magnetoelectric coupling. By applying out-of-plane DC magnetic fields, we can measure the induced polarization in ScAlN performing quasi-static charge measurements by an electrometer connected to the devices. This study extends our previous polymer-based magnetoelectric measurements on magnetostrictive substrate[24] to fully inorganic, silicon-compatible devices, confirming the feasibility of direct magnetoelectric detection in submicron structures and paving the way for energy efficient magnetic technology.

## 2. Results and discussion

### 2.1. Device fabrication

In this study, we explored the direct magnetoelectric coupling at the nanoscale in ScAlN/Ni thin-film composites by measuring the charge accumulation induced by variations in magnetization. The magnetoelectric heterostructure comprised a 100 nm-thick piezoelectric $Sc_{0.4}Al_{0.6}N$ (ScAlN) layer coupled with a magnetostrictive Nickel (Ni) layer, the thickness of which was either 100 nm or 200 nm. The ScAlN layer was deposited on a 50 nm-thick Ru layer, which served as the bottom electrode (BE) for electrical measurements. All constituent layers of the stack were deposited via sputtering onto a $Si/SiO_2$ (600 nm) substrate.

A schematic illustration of the nanoscale device used in this investigation is provided in **Figure 1**a. Spin-On-Carbon (SoC) was selected as the planarization material due to its low dielectric constant (~3.0), minimizing parasitic capacitance, and its low Young's modulus (14 GPa), which reduces undesired lateral clamping of the magnetoelectric stack.[25] The top electrode was formed by 20 nm Cr and 120 nm Au layer, but it is denoted as "Au (TE)" in the figures for simplicity.

Five different device geometries, with distinct aspect ratios, were investigated: 500 nm x 500 nm; 500 nm x 2 μm; 500 nm x 5 μm; 1 μm x 1 μm; and 1 μm x 2 μm, all with the same ScAlN and Ni layer thicknesses. As previously noted, to evaluate the magnetoelectric effect, we measured the charge accumulation in the ScAlN layer as a function of the magnetization of the Ni layer. The charge generated by the direct magnetoelectric effect $\Delta Q_{ME}$ scales linearly with the device area, according to the following expression:

$$\Delta Q_{ME} = A \cdot d_{33} \cdot \lambda \cdot Y \qquad (1)$$

where $A$ is the area of the device, $d_{33}$ is the piezoelectric coefficient of ScAlN, $\lambda$ is the is the magnetostriction coefficient of Ni, and $Y$ is the Young's modulus of ScAlN. This expression is valid under the assumption of full strain transfer within the composite. Furthermore, in the case of fully magnetized systems, $\lambda$ corresponds to the saturation magnetostriction of Ni $\lambda_S$.

Based on ideal material parameters, the estimated charge accumulation from Equation (1) for individual nanoscale devices was found to be of order of $10^{-17}$ C (0.01 fC), which is well below the experimental detection threshold. Therefore, in order to achieve detectable signal levels in the picocoulomb range, the effective device area was increased by fabricating arrays of devices connected in parallel. For devices with a 500 nm x 500 nm area, $10^6$ pillars were required; for larger device geometries, the number of pillars was proportionally reduced to maintain a constant total area and, consequently, a constant total capacitance.

Figure 1b presents a three-dimensional Atomic Force Microscopy (AFM) image showing the topography of multiple 1 μm x 1 μm devices after nanopillar and bottom electrode patterning. After the planarization step, top electrode lines were defined as perpendicular to the bottom electrode ones, as depicted in the microscope image in Figure 1c. This orthogonal layout was designed to minimize the overlap between the top and bottom electrodes, thereby reducing parasitic capacitance. Furthermore, cross-sectional analysis of the fabricated devices was conducted using a combination of Focused Ion Beam (FIB) milling and Scanning Electron Microscopy (SEM). Figure 1d displays an FIB cross-section of a 1 μm-wide device, while Figure 1e shows a set of three 500 nm-wide nanopillars connected by a shared Au top electrode line. As observed from these images, due to the ion beam etching process, the nanopillars acquire a slightly trapezoidal shape, leading to an effective area larger than the nominal planar design. This experimental area variation is taken into account in the following calculations.

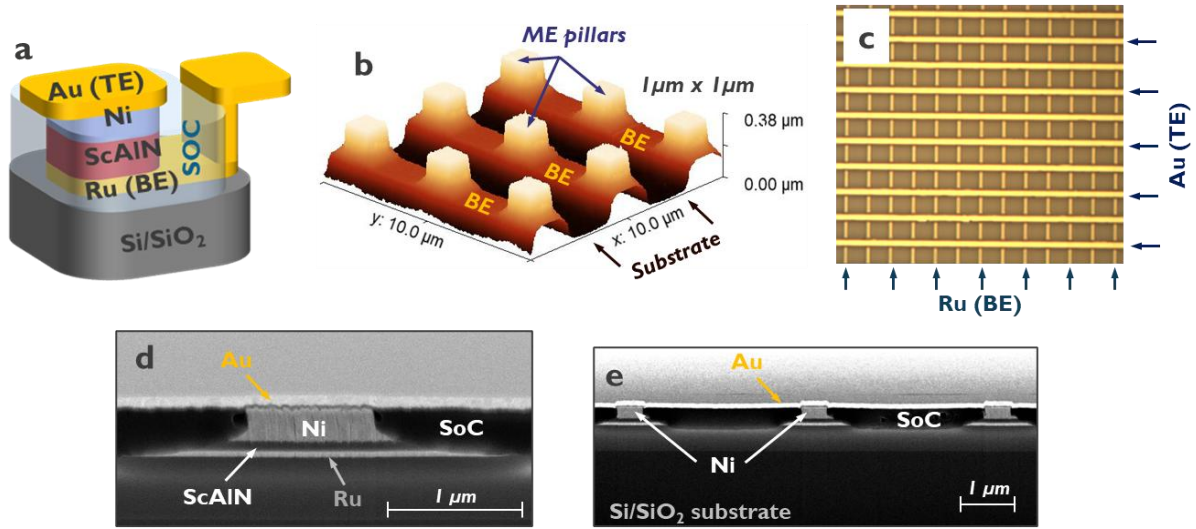

**Figure 1.** (a) Schematic representation of a single device fabricated by e-beam lithography on a Si/SiO2 substrate. (b) 3D-AFM image of 1 μm x 1 μm devices array with the magnetoelectric stack and the BE lines patterned by ion beam etching. (c) Microscope image showing the intersection between the TE lines (horizontal) and the BE lines (vertical). FIB image showing the cross-section of (d) one single 1 μm wide device and (e) 500 nm wide devices array post fabrication.

## 2.2. Magnetic characterization

The magnetization behavior of blanket Ni thin films (100 nm and 200 nm) fabricated on ScAlN was characterized by vibrating sample magnetometry (VSM) measurements, comparing the results with those of a bulk Ni foil. The measurements were performed by applying the magnetic field both parallel to the sample (In-Plane, IP) and perpendicular to the sample (Out-Of-Plane, OOP). The results of the magnetization-field (M–H) hysteresis loop in the IP and OOP directions are shown in Figure 2a and Figure 2b, respectively. It can be observed that all the samples show a saturation magnetization $M_S$ of approximately 470 kA/m, consistent with reported values for bulk Ni.[26] Furthermore, the Ni foil exhibits negligible coercivity both in the IP and OOP direction, indicative of soft magnetic behavior. On the other hand, the Ni films show an IP coercive field $H_{C,IP}$ of 12 mT and an OOP coercive field $H_{C,OOP}$ of 18 mT. This increase in coercivity is likely attributed to in-plane magnetic anisotropy induced by strain in the Ni film, associated with deposition on the underlying ScAlN layer.[27,28] Moreover, the IP saturation field increases with the thin film thickness, in particular Ni 200 nm > Ni 100 nm >

Ni foil. This trend is consistent with an increased magnetoelastic and shape anisotropy in thicker films, which require stronger fields to overcome.[29,30]

To characterize the Ni magnetic behavior at the nanoscale, single-nitrogen-vacancy (NV) center magnetometry inspections were performed on a single 1 µm x 1 µm device after the ion beam etching step, but prior deposition of the Au top electrode. This technique is used to quantitatively determine the absolute magnetic field projections along the parallel ($\mu_0 H_{NV\parallel}$) and perpendicular ($\mu_0 H_{NV\perp}$) directions to the NV axis, and it can be used to map the magnetic stray field generated by domain patterns in ferromagnets. A schematic representation of the NV-axis orientation used for the inspection is reported in Figure 2c. The maps of both magnetic field projections, and the resulting absolute field ($\mu_0 |\boldsymbol{H}|$) obtained for the Ni 1 µm x 1 µm device, are shown in Figure 2(d-f). The dark spots in the maps are measurements artefacts and appear where it is not possible to extract both NV electron spin resonance (ESR) frequencies from the optically detected magnetic resonance (ODMR) spectrum. This happens at small longitudinal fields due to resonance overlap, and at large perpendicular fields due to fluorescence quenching.[31] Prior to recording these maps, the sample was fully magnetized by applying a field of 500 mT in the OOP direction, which was subsequently removed. The inspection at remanence is performed at a small bias field of about 3 mT, aligned approximately along the in OOP direction.

Figure 2d and Figure 2e show the components of the magnetic stray field along the direction parallel to the NV axis (longitudinal field) and in the plane perpendicular to it (transversal field), respectively. Both images display multiple field maxima and minima, implying non-uniform magnetization in the Ni film. Furthermore, the longitudinal field distribution suggests a balanced presence of positive and negative stray fields, reflecting a multi-domain state in the magnetization. This interpretation is further supported by the absolute field map in Figure 2f, where local minima correspond to regions in which the longitudinal and transversal components both become small. These low-field regions are consistent with cancellation of stray fields from differently magnetized domains.

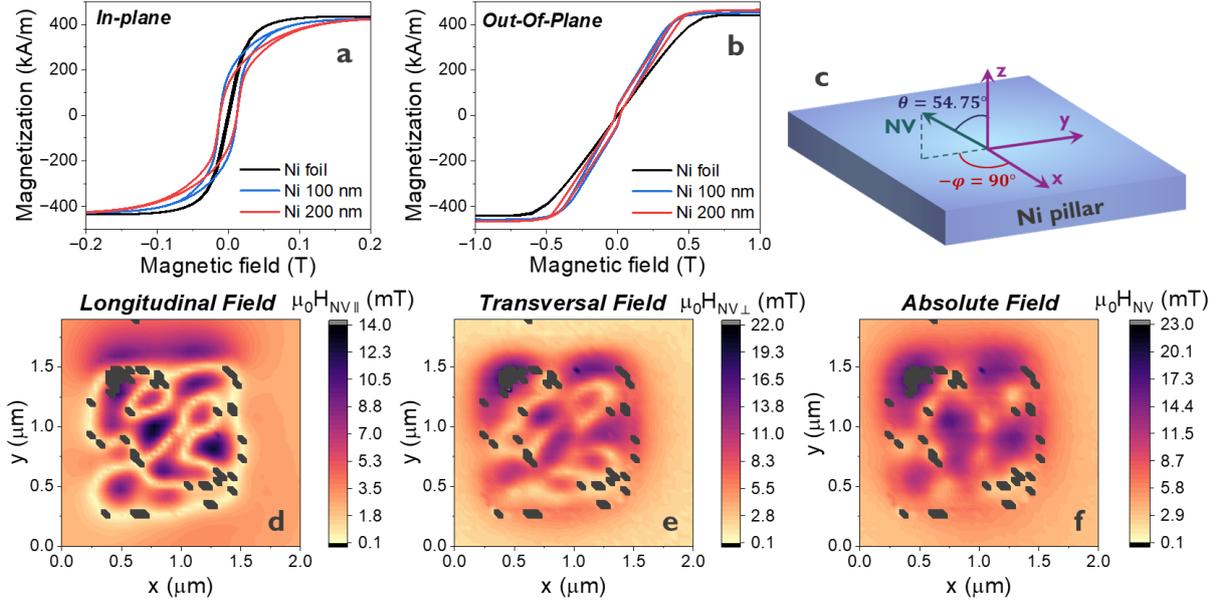

**Figure 2.** (a) In-Plane and (b) Out-of-plane ferromagnetic hysteresis loop (M-H) of blanket Ni measured by vibrating sample magnetometry (VSM); the data show a comparison between bulk Ni foil and thin film Ni (100 nm and 200 nm) fabricated on a ScAlN layer. (c) Schematic representation of single-nitrogen-vacancy (NV) center magnetometry technique used to study the Ni magnetic properties at the device level; the image shows the NV-axis orientation with respect to the Ni pillar device. 2D colormaps of the stray magnetic field at remanence of a single 1 μm x 1 μm device, showing the components along the (d) parallel (longitudinal field $\mu_0 H_{NV\parallel}$) and (e) perpendicular (transversal field $\mu_0 H_{NV\perp}$) directions of the NV axis, and (f) resulting absolute field $\mu_0 H_{NV}$.

## 2.3. Electrical characterization

The electrical behavior of the nanostructured devices was characterized through capacitance measurements using an LCR meter. The measured capacitance $C_{meas}$ was used to extract the effective dielectric constant of ScAlN at the device level, $\kappa_R^{dev}$, according to the following relation:

$$\kappa_R^{dev} = \frac{C_{meas} \cdot t}{\kappa_0 \cdot A_{FIB} \cdot n} \qquad (2)$$

where $t$ is the thickness of the ScAlN layer, $\kappa_0$ is the vacuum permittivity, $A_{FIB}$ is the actual contact area at the base of each nanopillar, determined from FIB cross-sections (see Figure 1d), and $n$ is the number of nanopillars in the array. The value obtained averaging the measurements on different devices was $\kappa_R^{dev} = 16.7 \pm 0.8$. The experimental error was calculated as the standard deviation across the measurements, and it is likely due to area variations between devices and to differences in the number of functional nanopillars within each array.

To provide a reference, this value was compared with the ScAlN dielectric constant $\kappa_R^{ref} = 16.0$, measured on unpatterned (blanket) films. Notably, the extracted device dielectric constant $\kappa_R^{dev}$ matches the blanket-film reference value $\kappa_R^{ref}$ within the experimental error. This consistency suggests that parasitic capacitances, such as from fringing fields or from the surrounding SoC layer, are negligible in these structures, supporting the use of the nanopillar array as a reliable platform for magnetoelectric investigations at the nanoscale.

Finally, leakage current measurements were performed on full arrays of devices with varying lateral dimensions. The results revealed leakage currents ranging from several hundred pA to a few nA, with a general trend of increasing leakage as the devices size decreases. This behaviour may be attributed to damage induced by ion beam etching, which tends to have a more pronounced effect on smaller structures. In particular, because of the higher number of repetitions, devices with smaller lateral dimensions (e.g., 500 nm × 500 nm) exhibit a larger total perimeter compared to larger devices (e.g., 1 μm × 2 μm), further contributing to the increased leakage. Moreover, partial overlap between the top and bottom electrodes within the planarization SoC layer was unavoidable, and this overlap becomes more pronounced as device dimensions decrease. While the low dielectric constant of the SoC layer limits its contribution to the overall capacitance, it may nonetheless influence the observed leakage behaviour.

### 2.4. Magnetoelectric measurements

To quantify the magnetoelectric response in nanoscale devices based on ScAlN/Ni composites, magnetic-field-induced charge accumulation $\Delta Q_{ME}$ was measured using an electrometer. A DC magnetic field was applied in the OOP direction across the entire device array using an electromagnet, sweeping from –1.0 T to 1.0 T. A schematic of the experimental setup is shown in **Figure 3**a.

The applied magnetic field H gradually magnetizes the Ni layer along the OOP direction, as already shown in the VSM image in Figure 2b. The magnetization varies within the field range of -0.45 T to 0.45 T, while saturation occurs beyond ±0.45 T, where magnetization and thus magnetostrictive strain remain constant. The strain ε generated in the Ni layer due to magnetostriction is transferred to the ScAlN layer, leading to charge accumulation $\Delta Q_{ME}$ in the piezoelectric material, which serves as a measure of the direct magnetoelectric response. The relationship between charge accumulation and the applied DC magnetic field for device arrays of various dimensions is presented in Figure 3b. All devices display charge variation within the same region of the Ni magnetization (–0.45 T to 0.45 T) and reach saturation above ±0.45 T. These findings align with the expected magnetoelectric behavior, that correlates the magnetoelectrically generated charges with the magnetization-induced strain.

The accumulated charge was used to estimate the equivalent open-circuit voltage $\Delta V_{ME}$, as previously reported,[24] by dividing $\Delta Q_{ME}$ by the experimentally determined array capacitance $C_{meas}$. These results are shown in Figure 3c. The measured $\Delta Q_{ME}$ and corresponding $\Delta V_{ME}$ ranged from $(0.24 \pm 0.06)\ pC$ and $(0.41 \pm 0.07)\ mV$ for the 500 nm x 500 nm device array, to $(0.57 \pm 0.09)\ pC$ and $(1.18 \pm 0.18)\ mV$ for the 1 µm x 1 µm and 500 nm x 2 µm device arrays, respectively. Measurement errors were derived from the standard deviation of the measurements performed within the saturation regions. These results are in agreement with finite element simulations by COMSOL Multiphysics, which predicted voltages in the same order of magnitude as the ones experimentally measured.

However, as stated in Equation (1), $\Delta Q_{ME}$ depends primarily on the device area, while $\Delta V_{ME}$ is determined by the ScAlN thickness $t$, as expressed by:

$$\Delta V_{ME} = t \cdot d_{33} \cdot \kappa_0^{-1} \cdot \kappa_R^{-1} \cdot \lambda \cdot Y \tag{3}$$

Given that the device arrays were designed to maintain a consistent overall area, and that the ScAlN thickness was uniform, the observed discrepancies in charge and voltage outputs for the different devices likely stem from variations in leakage currents or fabrication yield. Notably, the 500 nm x 500 nm array exhibited relatively lower performance, possibly due to the increased leakage current in smaller dimension devices, as explained in the previous section. Overall, these results confirm that a measurable magnetoelectric response can be achieved in nanoscale thin-film composites, despite substrate clamping effects. The resulting charge accumulation and open-circuit voltages in the millivolt range highlight the potential of these structures for spintronic applications.

To gain deeper insight into the magnetoelectric response of nanoscale thin-film composites, additional measurements were conducted. The first analysis focused on the influence of the thickness ratio between the magnetostrictive (Ni) and piezoelectric (ScAlN) layers on the magnitude of the magnetoelectrically generated charge and voltage. In this study, the ScAlN thickness was held constant at 100 nm, while the Ni thickness was varied between 100 nm (thickness ratio = 1) and 200 nm (thickness ratio = 2). The corresponding experimental results, obtained from 1 μm x 2 μm device arrays, are presented in Figure 3d. As shown, increasing the Ni thickness leads to a significant enhancement in the magnetoelectric response. An approximate 30% increase is observed in both the generated charges $\Delta Q_{ME}$ and voltages $\Delta V_{ME}$ when the thickness of the magnetostrictive layer is doubled. This enhancement is attributed to a more efficient strain transfer from the magnetostrictive to the piezoelectric layer. Thicker magnetostrictive films generate greater strain under magnetic excitation due to their reduced susceptibility to substrate-induced clamping, enabling more effective mechanical coupling with the piezoelectric layer. These findings are consistent with previous studies reporting that increasing the thickness of the magnetostrictive layer, or more generally the magnetostrictive-to-piezoelectric thickness ratio, can significantly enhance the magnetoelectric effect in composite structures.[5,32,33]

Finally, magnetoelectric measurements were performed on the 1 μm x 1 μm device array under both increasing and decreasing OOP magnetic field sweeps, as shown in Figure 3e. The corresponding measurement traces, labeled "forward" (increasing field) and "backward" (decreasing field), exhibit a largely symmetric magnetoelectric response, consistent with the expected linear behavior. However, a systematic offset of approximately 100 mT is observed between the forward and backward magnetic field sweeps. Specifically, the forward sweep is shifted by ~50 mT toward positive magnetic fields, while the backward sweep is shifted by ~50 mT toward negative fields. This hysteretic shift is consistent with the OOP M-H hysteresis behavior observed in thin Ni films deposited on ScAlN. The observed increase in coercivity, from 18 mT in blanket films to approximately 50 mT at the device level, is likely due to the geometry of the fabricated structures. In particular, the Ni layer in these nanopillars has a thickness of 200 nm, which is comparable to its lateral dimension (1 μm), leading to significant shape anisotropy effects. Similar phenomena have been reported in patterned ferromagnetic nanostructures, where geometric confinement and shape-induced anisotropy significantly influence the magnetic response.[34]

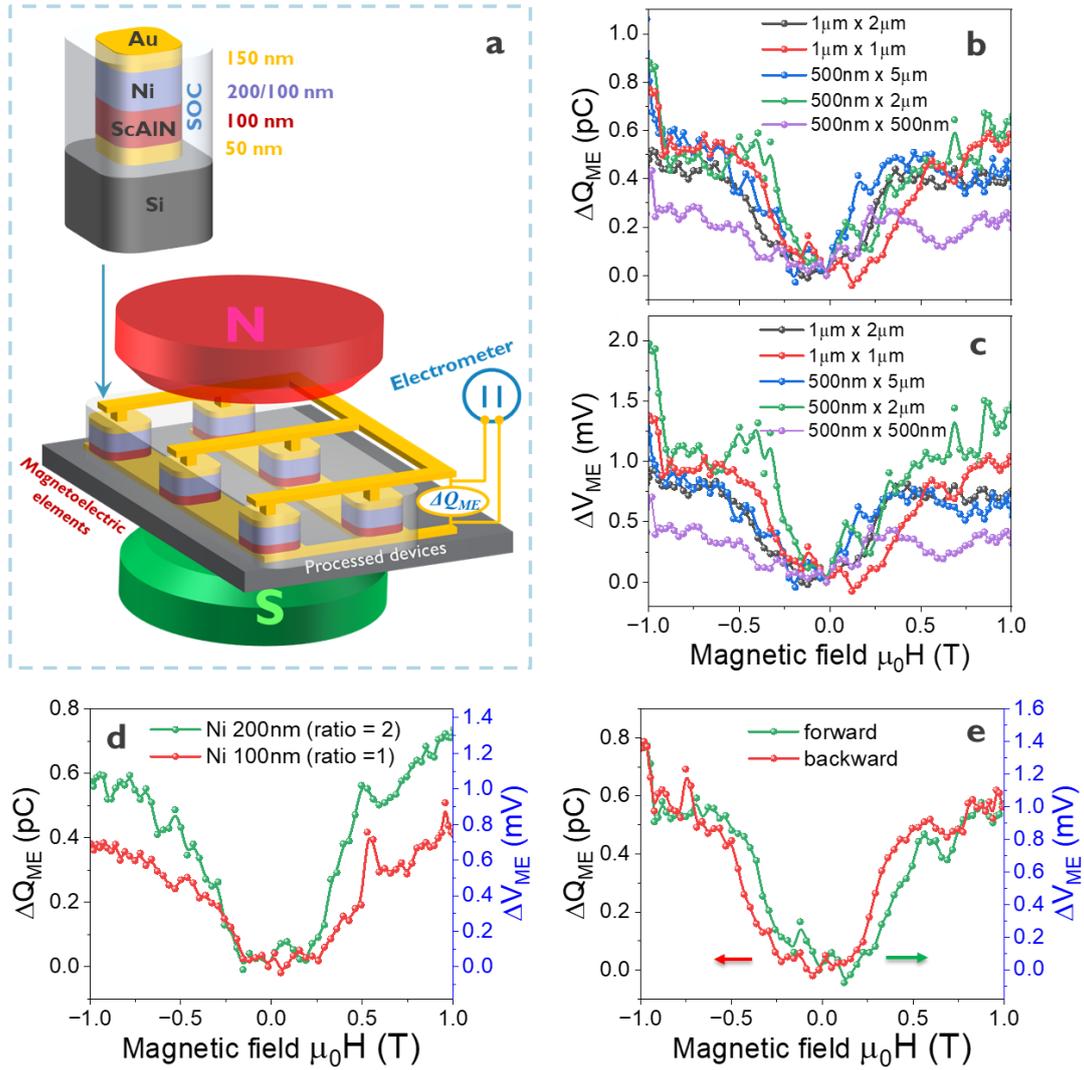

**Figure 3.** (a) Schematic representation of magnetoelectric measurements setup for magnetic-field induced charge accumulation on nanoscale devices based on ScAlN/Ni composites. Magnetoelectrically generated (b) charge accumulation $\Delta Q_{ME}$ and (c) open-circuit voltage $\Delta V_{ME}$ for device arrays with different nanopillars dimensions. (d) Dependence of the magnetoelectrically generated charges $\Delta Q_{ME}$ and voltages $\Delta V_{ME}$ in the 1 μm x 2 μm device array on the Ni layer thickness (i.e. on the magnetostrictive vs piezoelectric thicknesses ratio). (e) Magnetoelectrically generated charges $\Delta Q_{ME}$ and voltages $\Delta V_{ME}$ measurements performed on the 1 μm x 1 μm device array with increasing (forward) and decreasing (backward) magnetic field.

## 3. Conclusion

In conclusion, we experimentally investigated the direct magnetoelectric effect in nanoscale ScAlN/Ni thin-film composites. Arrays of devices with lateral dimensions down to 500 nm x 500 nm were fabricated using electron beam lithography and ion beam etching, with structural characterization confirmed via FIB cross-sectional imaging.

Ferromagnetic characterization of Ni thin films deposited on ScAlN was conducted using VSM, revealing an increase of the coercivity and IP saturation field in comparison to bulk Ni foils. These results highlight the influence of strain-induced anisotropy on the M-H hysteresis behavior. Additionally, NV center magnetometry was employed to investigate the magnetic configuration of the Ni at the device level, demonstrating the formation of a complex domain state at remanence. Finally, capacitance measurements on the device arrays yielded a dielectric constant of $\kappa_R^{\text{dev}} = 16.7 \pm 0.8$, in agreement with the reference value from blanket ScAlN films ($\kappa_R^{ref} = 16.0$).

The direct magnetoelectric effect was probed through quasi-static charge measurements $\Delta Q_{ME}$ under an applied DC magnetic field ranging from –1.0 T to +1.0 T. The corresponding open-circuit voltage was obtained by normalizing the measured charge to the device capacitance, yielding values up to $\Delta V_{ME} = (1.17 \pm 0.18)$ mV. Variation of the Ni thickness revealed an enhancement in magnetoelectric coupling with increasing magnetostrictive-to-piezoelectric thickness ratio, in agreement with theoretical expectations for strain-mediated composites. Additionally, magnetoelectric voltage hysteresis in 1 µm x 1 µm devices with 200 nm Ni showed a coercive shift of approximately 100 mT between forward and reverse magnetic field sweeps. This hysteretic behaviour is in qualitative agreement with the blanket M-H hysteresis loop, with an increased coercivity at the device level attributed to shape anisotropy.

These findings represent the first demonstration of quasi-static direct magnetoelectric coupling in nanoscale heterostructures fabricated on a rigid silicon substrate. They confirm the feasibility of strain-mediated magnetoelectric detection in submicron structures and open promising avenues for scalable, energy-efficient magnetic memory and sensing technologies.

## 4. Experimental Section

*Device fabrication procedure*: Device fabrication was carried out using electron-beam (EB) lithography using a F700S EB system. It consisted of five primary steps: (i) patterning of the magnetoelectric stack into nanopillars via ion beam etching; (ii) patterning of Ru bottom electrode lines by ion beam etching; (iii) planarization of the nanopillars using Spin-on-Carbon (SoC); (iv) patterning of Chromium (Cr, 20 nm) / Gold (Au, 120 nm) top electrode lines via lift-off; and (v) formation of Au contact pads to the top and bottom electrodes, also by lift-off. The planarization step was performed spin coating a Hexamethyldisilazane (HDMS) adhesion promoter at 2000 rpm for 1 minute and subsequently a Nissan GF-345SN SoC at 1000 rpm for 1 minute. The two layers were then baked at 250 °C for 1 minute in $N_2$ atmosphere. The recession till the device level was performed by reactive ion etching (RIE) in a $N_2$/ $H_2$ atmosphere using an Oxford Plasmalab 100 etching system.

*Device imaging*: The 3D images of the devices post ion beam etching were obtained by atomic force microscopy (AFM) using a Bruker Dimension Edge instrument. The device cross-section was imaged using a dual beam Focused Ion Beam (FIB) / Scanning Electron Microscope (SEM) Helios460HP.

*Magnetic measurements*: Ferromagnetic properties of the Ni layer were investigated by magnetization–magnetic field (M–H) hysteresis loop measurements employing a MicroSense EV11 vibrating sample magnetometer (VSM). The out-of-plane magnetic field was applied within ±1.0 T, while the in-plane magnetic field was applied within ±0.2 T. Scanning Single Nitrogen Vacancy (NV) Center Magnetometry[35] measurements were performed using a commercial setup (ProteusQ, Qnami). The diamond probe (MX series) was set at a distance of 100 nm from the sample surface to reduce fluorescence quenching, allowing the extraction of the NV ESR frequencies using continuous wave optically detected magnetic resonance (cw-ODMR). From the ESR frequencies, the longitudinal and transverse magnetic field components are calculated using the relations provided by Van Der Sar et al.,[36] assuming a zero-field splitting of the NV electronic ground state $D = 2.869$ GHz and the electron gyromagnetic ratio $\gamma = 28$ GHz/T. The NV orientation is roughly at an angle $\theta = 54.75°$ from the z axis and $\varphi = 270°$ from the x axis. Prior to the measurement, the magnetic film is fully saturated by applying an OOP magnetic field of 500 mT using an electromagnet. The sample is then inserted into the scanning NV microscope to perform the characterization of the Ni remanence state in a small bias field of about 3 mT in OOP direction.

*Device characterization*: Dielectric properties of ScAlN at the device level were characterized by capacitance measurements conducted with an E4980A precision LCR meter at a frequency of 10 kHz and an AC voltage of 100 mV (RMS). To characterize the magnetoelectric coupling, out-of-plane magnetic field sweeps between -1.0 T and 1.0 T were applied to the devices. The generated charge due to the magnetoelectric effect was simultaneously measured using a Keithley 6517B Electrometer.[24]

**Acknowledgements**

This work has been supported by IMEC's industrial affiliate program on exploratory logic. F.L. and E.V.M. acknowledge support by the Research Foundation Flanders (FWO) through PhD fellowships under grant agreements No. 1183722N and No. 1SH4Q24N, respectively. F.C., C.A., E.S., P.P. and M.W. acknowledge the financial support received from the Horizon Europe research and innovation program within the project MandMEMS (Grant Agreement No. 101070536). Finally, F.L., E.V.M., S.D.G., F.C. and C.A. acknowledge the help from IMEC's e-beam lithography team for the sample preparation, and the IMEC's MCA department for the FIB imaging.

**Conflict of Interest**

The authors declare no conflict of interest.

**Data availability**

The data that support the findings of this study are available from the corresponding author upon reasonable request.